\documentclass[a4paper,conference]{IEEEtran}

\usepackage[edges]{forest}
\usepackage{tikz}
\usepackage{etoolbox}
\usepackage{multirow}
\usepackage{makecell}
\usepackage{amsmath,amsfonts,amssymb}
\usepackage{graphicx}
\usepackage{setspace}
\usepackage{tocloft}
\usepackage{lineno}
\usepackage{cite}
\usepackage{color}
\usepackage{comment}
\usepackage{amsmath,amssymb}
\usepackage[utf8]{inputenc}
\usepackage{booktabs}
\usepackage{floatrow}
\usepackage{subcaption}
\usepackage{algorithm}
\usepackage{algorithmic}
\usepackage{xhfill}
\usepackage[belowskip=-2pt,aboveskip=1pt]{caption}
\usepackage{titlesec}
\titlespacing*{\section}{0pt}{0.6\baselineskip}{0.6\baselineskip}
\titlespacing*{\subsection}{0pt}{0.5\baselineskip}{0.5\baselineskip}
\titlespacing*{\subsubsection}{0pt}{0.5\baselineskip}{0.5\baselineskip}
\titlespacing{\section}{0pt}{2\parskip}{-2\parskip}
\titlespacing*{\section}{0pt}{2\parskip}{-2\parskip}
\titlespacing{\subsection}{0pt}{2\parskip}{-2\parskip}
\titlespacing{\subsubsection}{0pt}{25\parskip}{-25\parskip}

\ifCLASSINFOpdf

\else

\fi

\begin{document}
%
\title{Offset Curves Loss for Imbalanced Problem \\ in Medical Segmentation}

\author{\IEEEauthorblockN{Ngan Le, Trung Le, Kashu Yamazaki, Toan Bui, Khoa Luu, Marios Savides}
\IEEEauthorblockA{University of Arkansas, Fayetteville, Arkansas USA 72703\\
Email: \{thile\}@uark.edu }
\IEEEauthorblockA{University of Carnegie Mellon University, Pittsburgh, PA 15213\\
Email: thihoanl@andrew.cmu.edu, msavvid@ri.cmu.edu}
}


%


\maketitle

\begin{abstract}
Medical image segmentation has played an important role in medical analysis and widely developed for many clinical applications. Deep learning-based approaches have achieved high performance in semantic segmentation but they are limited to pixel-wise setting and imbalanced classes data problem. In this paper, we tackle those limitations by developing a new deep learning-based model which takes into account both \textbf{higher feature level} i.e. region inside contour, \textbf{intermediate feature level} i.e. offset curves around the contour and \textbf{lower feature level} i.e. contour. Our proposed Offset Curves (OsC) loss consists of three main fitting terms. The first fitting term focuses on pixel-wise level segmentation whereas the second fitting term acts as attention model which pays attention to the area around the boundaries (offset curves). The third terms plays a role as regularization term which takes the length of boundaries into account. We evaluate our proposed OsC loss on both 2D network and 3D network. Two common medical datasets, i.e. retina DRIVE and brain tumor BRATS 2018  datasets are used to benchmark our proposed loss performance. The experiments have shown that our proposed OsC loss function outperforms other mainstream loss functions such as Cross-Entropy, Dice, Focal on the most common segmentation networks Unet, FCN.
\end{abstract}

\IEEEpeerreviewmaketitle

\section{Introduction}
\label{sec:intro}

\begin{figure*}[!t]
	\centering
	\includegraphics[width=18cm]{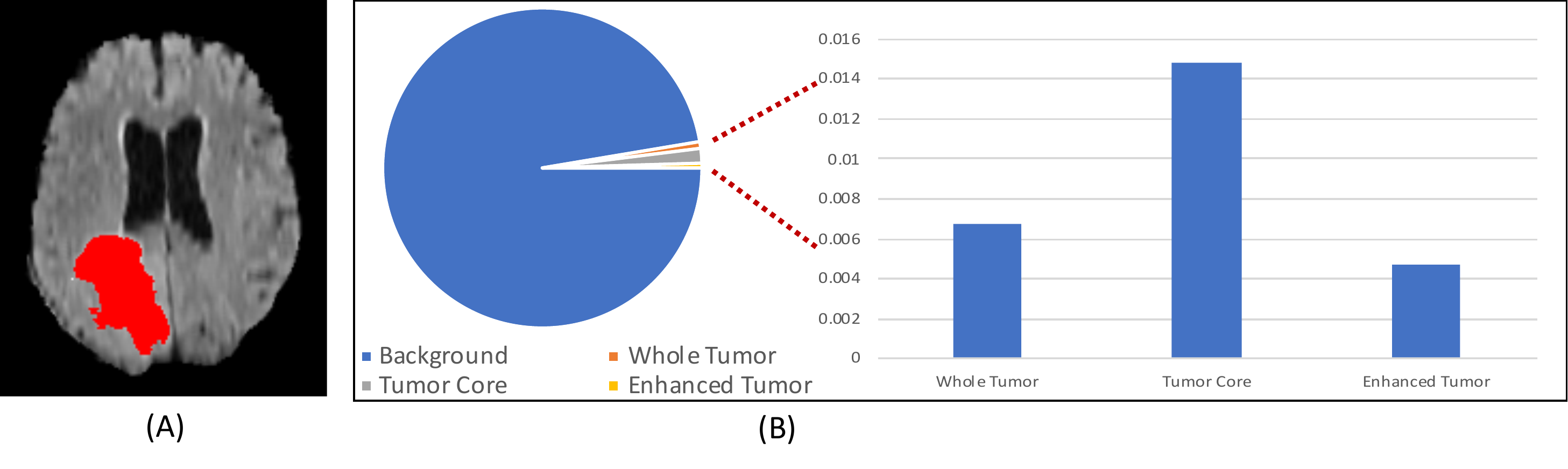}
	\caption{(A): an MRI slice visualizing a brain tumor. No. of brain tumor pixels: 1,410 (2.4\%) and no. of brain pixels: 57,600(97.56\%). (B): statistical information about the ratio between different classes in BRATS2018}
	\label{fig:imbalance}
\end{figure*}

Class imbalance is a common problem in medical segmentation when  treating it as a classification problem, where for every pixel in the image, we try to predict what it is. Take brain tumor as an instance, the number of pixels belonging to foreground class (tumor) is quite small compared to the number of pixels belonging to background class (non-tumor).  Fig. \ref{fig:imbalance}(A) shows a middle slice of a brain tumor MRI where the tumor occupies 2.4\% of entire slide where Fig. \ref{fig:imbalance}(B) shows the statistical information about the ratio between different classes in BRATS2018 \cite{Brats} dataset. This imbalanced data problem makes the requirement of high accurately segmenting brain tumor become more difficult.

Class imbalance has been studied thoroughly over the last decades using either traditional machine learning models, i.e. non-deep learning or advanced deep learning techniques. Despite recent advances in deep learning, along with its increasing popularity, very few works in the area of deep learning with class imbalance exist. The previous deep learning-based approaches, which address the class imbalance problem, can be mainly divided into three groups: data-level methods, algorithm-level methods and hybrid-level methods as follows: (i) Data level methods: aims at altering the training data distribution by either adding more samples into minority class or removing samples from the majority class to compensate for imbalanced distribution between classes; (ii)  Algorithm level methods: aim at making a modification to the conventional learning algorithms to reduce imbalanced between classes; (iii) Hybrid methods: Those methods are a combination of the merits of both data level and algorithm level strategies.

\begin{figure*}	
\centering
\begin{forest}
[Deep Learning to Class Imbalance
[Data-Level 
[\rotatebox{85}{Under-sampling \cite{Lee_2016}}
]
[\rotatebox{85}{Over-sampling \cite{Masko_2015}}
]
[\rotatebox{85}{Dynamic-sampling\cite{Samira_2018}}
]
]
[Algorithm-Level
[ Loss Function
[\rotatebox{85}{MFE/MSFE\cite{Wang_2016}}
]
[\rotatebox{85}{Focal Loss\cite{Focalloss}}
]
[\rotatebox{85}{Boundary Loss \cite{MIDL2019_loss}}
]
[\rotatebox{85}{Class Rectification \cite{Qi_2017}}
]
]
[Cost-Sentivity
[\rotatebox{85}{CoSen CNN\cite{Khan_2017}}]
[\rotatebox{85}{CSDBN-DE\cite{Khan_2017}}]
[\rotatebox{85}{Q-learnig-based\cite{Lin_2019}}]
]	
]
[Hybrid-Level
[\rotatebox{85}{Large Margin Local \cite{Huang_2016}}
]
[\rotatebox{85}{Deep Over Sampling \cite{Ando_2017}}
]
]
]
\end{forest}
\caption{Summary of Deep Learning approaches to imbalanced-class data problem}
\label{fig:fig_summary}
\end{figure*}
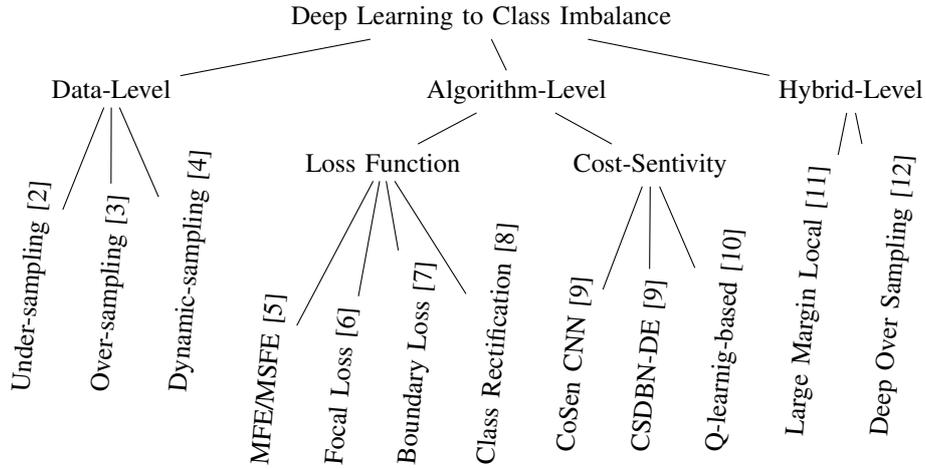

The literature review on imbalanced-class data problem is summarized as in Fig.\ref{fig:fig_summary}. \textit{Our proposed DL architecture with OsC loss for medical segmentation belongs to the second category} where we pay attention on proposing an effective loss for imbalanced problem in medical segmentation.

\begin{figure*}[!t]
	\centering
	\includegraphics[width=18cm]{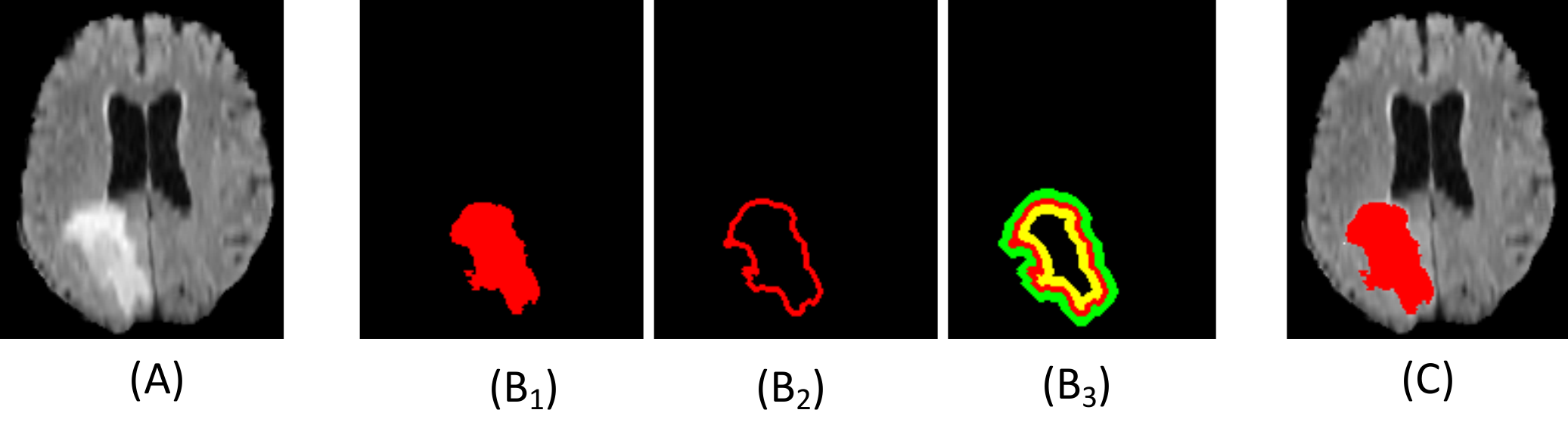}
	\caption{Given a image with imbalanced-class data ($A$), the proposed OsC loss focuses on three different levels of data: higher level feature - region($B_1$), lower level feature - contour ($B_2$) and intermediate feature level - offset curves ($B_3$). The final segmentation is a combination of three different feature levels and shown in ($C$)}
	\label{fig:loss}
\end{figure*}

\textbf{Our motivations and contribution:}
Medical image segmentation has been widely studied and developed for refinement of clinical analysis and applications. DL-based approaches \cite{Ronneberger_2015, Unet, 3DUnet, Milletari_2016, 3D_ESPNet_2018, S3D_UNet_2018} have achieved great success in both 2D and 3D (volumetric) data. Most DL network segmentation have made use of common loss functions e.g., Cross-Entropy, Dice \cite{Milletari_2016}, and the recent Focal \cite{Focalloss}. These losses are based on summations over the segmentation regions and are restricted to pixel-wise setting. \textit{Not only pixel-wise sensitivity, these losses do not take geometrical information into account as well as are limited to imbalanced-class data}. Furthermore class imbalance is naturally existing in the medical imaging segmentation problem which is considered as pixel level, i.e. each pixel is classified as tumor/not tumor (brain tumor) or foreground (blood vessel) or background (retina). In most applications, the number of pixels in each class are unbalanced as shown in Fig.\ref{fig:imbalance}. \textit{Our work aims at tackling those limitations by developing a new loss which takes into account both global and local information during learning including (i) lower feature level i.e contour; (ii) intermediate feature level i.e. narrow band around the contour; (iii) higher feature level i.e. area inside the contour} as illustrated in Fig.\ref{fig:loss}. That means our loss function not only pays attention at region information but also focuses on support pixels at two side of the boundary under the offset curves and the contour itself. Our proposed OsC loss is based of offset curves theory and active contours energy minimization. Thus, we first describe the theoretical framework of parallel curves and surfaces which will be detailed in Sec.\ref{subsec:offsetcurves}, we then details active contour with its successfully implementation, named variational level set in Sec.\ref{subsec:levelset}.

\section{Related works}
\label{sec:relatedwork}
\subsection{Variational Level Set (VLS)}
\label{subsec:levelset}
Segmentation by deformable models was first introduced by \cite{Kass1988}. From an initial location, these models deform according to an iterative evolution procedure until they fit structures of interest. Active Contour (AC) based on variational models and partial differential equations (PDEs), can be considered as one of the most widely used deformable models-based approaches in medical image segmentation. There are two main approaches in AC: explicit (i.e. snakes) and implicit (i.e Variational Level Set (VLS)). Snakes explicitly move predefined snake points based on an energy minimization scheme and unable to reach boundaries if their initial location is far from them whereas VLS approaches inspired by the Mumford-Shah functional \cite{Mumford_1989} move contours implicitly as a particular level of a function.
From the early work of using region terms in the evolution
of parametric snakes \cite{Song_1996}, region competition method \cite{Ivins_1995} to the recent works \cite{Li2011, Le_2016lip},  many approaches have dealt with region-based approaches using the variational level set (VLS) framework. VLS based or implicit AC is considered as a region-based deformable model which segment an image according to statistical data computed over the entire regions, i.e. the object of interest and the background. Even most VLS-based segmentation approaches are based on an assumption that image partitions should be uniform in terms of intensity, whether prior knowledge on the distribution of pixel intensities is available or not, VLS-based models have provided more flexibility and convenience for the implementation of active contours. Thus, they have been used in a variety of image processing and computer vision tasks. 


One of the most popular region based AC models is proposed by \cite{chan2001active}(CV). The CV-model has successfully segmented an image into two regions, that each region has a distinct mean of pixel intensity, by minimizing the energy functional. The  energy functional is defined based on the difference of image features, such as color and texture, between foreground and background. The fitting term or energy term in CV-model is defined by the inside contour energy and outside contour energy. The regularization term is defined by the length of the contour and the size of area inside the contour.

\subsection{Offset Curves Analysis}
\label{subsec:offsetcurves}
\begin{figure*}[!h]
    \centering
    \includegraphics[width= 13cm]{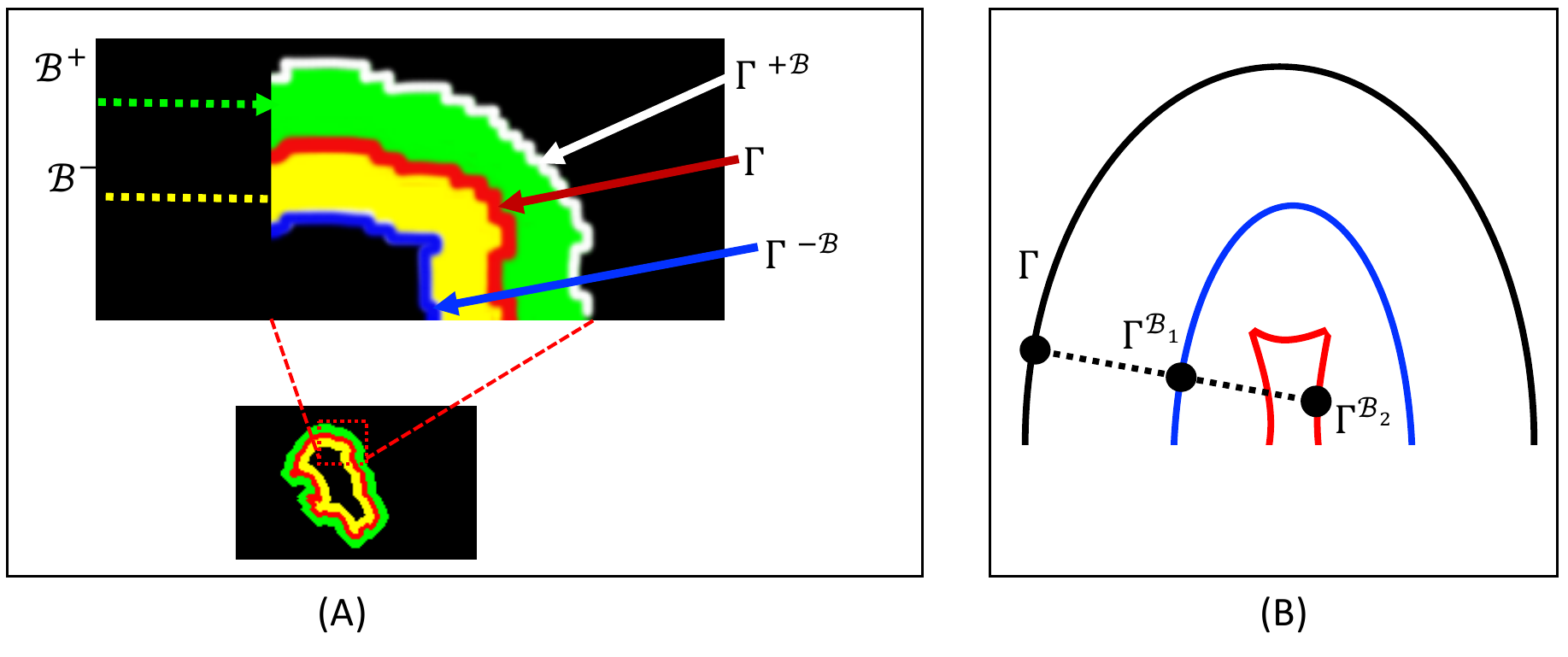}
    \caption{(A): Illustration of inner band $\mathcal{B}^{-}$ and outer bands $\mathcal{B}^{+}$ of a contour $\Gamma$ ; (B): Illustration of parallel curves theory with the main curve (black curve $\Gamma$) and its two parallel curves. Small translation $\mathcal{B}_1$ yields regular curve (blue curve) whereas large translation $\mathcal{B}_2$ yields a curve with singularities (red curve). The dashed line links corresponding points on parallel curves.}
    \label{fig:narrow}
\end{figure*}

The theoretical background of offset curves is based on the theory of parallel curves and surfaces \cite{gray2006modern, narrow_2}. An illustration of offset curve theory is given in Fig. \ref{fig:narrow}. In Fig.\ref{fig:narrow}(A), the curve $\Gamma$, where $\Gamma: \Omega \rightarrow \mathbb{R}^2 $ is called a parallel curve of $\Gamma^{\mathcal{B}}$ (either outer curve $\Gamma^{\mathcal{+B}}$ or inner curve $\Gamma^{\mathcal{-B}}$)  if its position vector $c^\mathcal{B}$ satisfies: 
\begin{equation}
    c^{\mathcal{B}}(z) =  c(z)  + \mathcal{B}n(z)
\label{eq:curve}
\end{equation}
where $z \rightarrow c(z) = [x(z), y(z)]$, $x$ and $y$ are continuously differentiable with respect to parameter $z$ and $\Omega \in [0,1]$. $\mathcal{B}$ is the amount of translation, and $n$ in the inward unit normal of $\Gamma$. Based on this equation, the inner band $\mathcal{B}^{-}$ and outer band $\mathcal{B}^{+}$ are bounded by parallel curves $\Gamma^{\mathcal{+B}}$ and $\Gamma^{\mathcal{-B}}$. This implies that both curves are continuously differentiable and do not exhibit singularities. Fig.\ref{fig:narrow}(B) shows a case where band width (translation) $B_1$ is  smaller than the curve’s radius of curvature whereas $B_2$ is larger than the curve’s radius of curvature. An important property resulting from the definition of the Eq.\ref{eq:curve} is that the velocity vector of parallel curves depends on the curvature of $\Gamma$. That means, the velocity vector of curve $\Gamma^{\mathcal{B}}$ is expressed as a function of the velocity vector,  curvature and normal of $\Gamma$. Set $n(z) = -\alpha c(z)$, we have 
\begin{equation}
\begin{split}
    c^{\mathcal{B}}(z)  =   c(z)  + \mathcal{B}n(z)= (1 - \alpha\mathcal{B})c(z)
\end{split}
\end{equation}
That equation provides the length element of inner parallel curve: $l^{\mathcal{B}} = ||c^{\mathcal{B}}(z)||= l^{\mathcal{B}}(1 - \alpha\mathcal{B})$
This is also a result in parallel curve theory in \cite{Elber_offset_1997}. Because the length $l^{\mathcal{B}}$ is also positive, the band width should not exceed the radius of curvature it is expressed as $\frac{-1}{\mathcal{B}} < \alpha < \frac{1}{\mathcal{B}}$. Is this constraint satisfies, the curves $\Gamma^{+\mathcal{B}}$ and $\Gamma^{-\mathcal{B}}$ are simple and regular.

\subsection{Loss function}
To train a Deep Neural Networks (DNNs), the loss function, which is known as cost function, plays a significant role. Loss function is to measure the average (expected) divergence between the output of the network i.e. predicted output ($P$) and the actual function, groundtruth ($T$) being approximated, over the entire domain of the input, sized $m \times n$; $i$ indexes each pixel value in image spatial space $N$; the label of each class is written as $l$ in $C$ classes. Herein, we briefly review some common loss functions. 

\textbf{Cross Entropy loss:} is a widely used pixel-wise measure to evaluate the performance of classification or segmentation model. For binary segmentation, CE loss is expressed as Binary-CE (BCE) loss function as follows:
\begin{equation}
    BCE = -\frac{1}{N}\sum_{i=1}^{N}{[T_iln(P_i) + (1 - T_i)ln(1-P_i)] }
\end{equation}
The standard CE loss has well-known drawbacks in the context of highly unbalanced problems.

\textbf{Dice loss:} measures the degree of overlapping between the reference and segmentation. Dice loss  \cite{Milletari_2016} is come from Dice score which was used to evaluate the segmentation performance as defined as:
\begin{equation}
    Dice = 1 - 2\frac{\sum_i^N{T_i P_i}}{\sum_i^N{T_i+P_i}} = 2\frac{T \cap P}{T \cup P}
\end{equation}
Even though Dice loss have been successful in image segmentation, it is still pixel-wise loss and has similar limitations as CE loss. Despite Dice loss improvements over CE loss, Dice losses may undergo difficulties when dealing with very small structures \cite{Sudre_2017} as misclassifying a few pixels can lead to a large decrease of the coefficient.

\textbf{Focal Loss (FL):} FL \cite{Focalloss} is a modified version of CE loss. It is to balances between easy and hard samples and is defined as follows:
\begin{equation}
\begin{split}
    FL = \frac{\alpha_i}{N}\sum_{i=1}^{N}{((1-P_i)^\gamma  T_i ln(P_i)   +P_i^\gamma ( 1 - T_i) ln(1- P_i))}
\end{split}
\end{equation}

In Focal loss, the loss for confidently correctly classified labels is scaled down, so that the network focuses more on incorrect and low confidence labels than on increasing its confidence in the already correct labels.  The loss focuses more on less accurate labels than the logarithmic loss when $\gamma > 1$

\section{Our Proposed Model}

\begin{figure*}[!t]
	\centering
	\includegraphics[width=18cm]{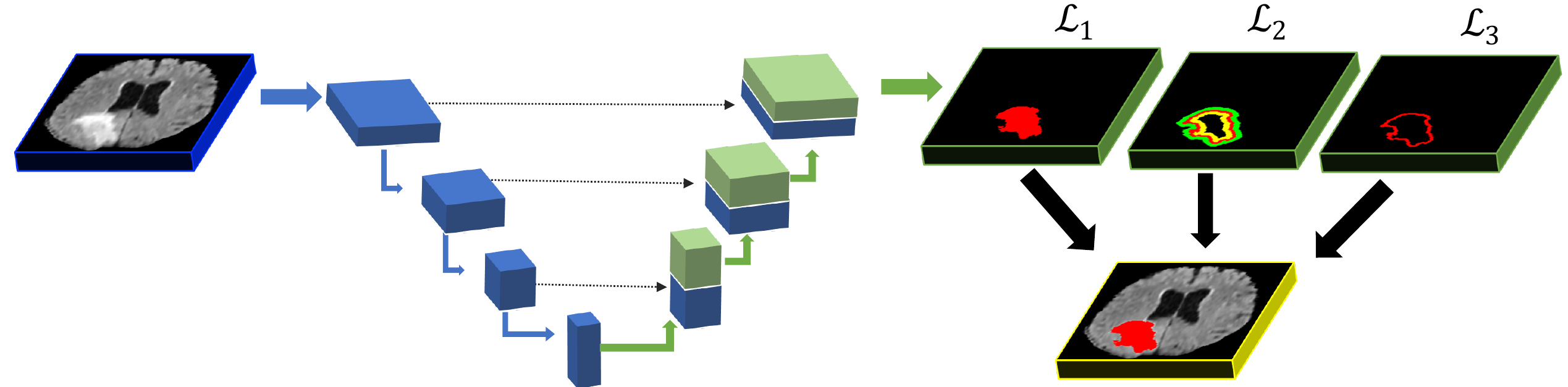}
	\caption{Illustration of our proposed OsC loss with Unet architecture. The loss between output from the network and the target contains three terms $\mathcal{L} = \mathcal{L}_1 + \mathcal{L}_2 + \mathcal{L}_3 $ where the first term $\mathcal{L}_1$ is region-based loss, the second term $\mathcal{L}_2$ is OsC loss  and the third term $\mathcal{L}_3$ is edge-based loss.}
	\label{fig:flowchart}
\end{figure*}

\subsection{Proposed OsC Loss}
Our proposed loss is motivated by the minimization problem of CV's model \cite{chan2001active} to efficiently find a contour by minimizing an energy functional. To address the limitations of CV's model (as in Sec. \ref{subsec:levelset}), we conduct an attention model to focus on parallel curves of the contour. In the following equations, ground truth and predicted output are denoted as $\textbf{T}$ and $\textbf{P}$, where $\textbf{T}, \textbf{P} \in [0,1]^{H\times W}$ and $H$ and $W$ are the height and weight of $\textbf{T}$. Our proposed OsC loss contains three fitting terms corresponding to higher feature loss, intermediate feature loss and lower feature loss as follows:

\textbf{Higher feature loss}: for a region segmentation of $K$ classes, the higher feature loss branch is computed as:
\begin{equation}
\mathcal{L}_{1} = -\sum_{c=1}^K{\textbf{T}_o^c log \textbf{P}_o^c} 
\end{equation}
where $\textbf{T}_o^c$ is binary indicator (0 or 1) if class label 'c' is the correct classification for observation 'o' and $\textbf{P}_o^c$ is predicted probability observation 'o' is of class 'c'.\\

\textbf{Intermediate feature loss}: This intermediate feature loss $\mathcal{L_2}$ is defined as the loss around the contour $\Gamma$. According to Sec.\ref{subsec:offsetcurves}, $\mathcal{B^{-}}$ and $\mathcal{B^{+}}$ are the inner band domain the outer band domain. Clearly, $\mathcal{B^{-}} \in \omega_I$ and  $\mathcal{B^{+}} \in \omega_O$ as shown in Fig.\ref{fig:narrow}(A). The $\mathcal{L}_2$ is computed inside the area $\mathcal{B} = [\mathcal{B^{-}}, \mathcal{B^{+}}]$ and defined:
\begin{equation}
\small
\begin{split}
\centering
    \mathcal{L}_{2} = &  \lambda_1\sum_{x,y\in\mathcal{B}}{|\textbf{P}(x,y)-b^{-}|^{2}H(\phi(x,y))} \\
 & + \lambda_2 \sum_{x,y\in\mathcal{B}}{|\textbf{P}(x,y)-b^{+}|^{2}(1-H(\phi(x,y))) } \\
\end{split}    
\label{eq:nbloss}
\end{equation}

where $b^{-}$ and $b^{+}$ are intensity descriptors of $\mathcal{B}^{-}$ and $\mathcal{B}^{+}$, respectively.
\begin{equation}
\begin{split}
b^{-} &  = \frac{\sum_{x,y\in\mathcal{B}}{\textbf{T}(x, y)\mathcal{H}(\phi(x,y))}}{\sum_{x,y\in\mathcal{B}}\mathcal{H}(\phi(x,y))} \text{ and } \\ 
b^{+} &  = \frac{\sum_{x,y\in\mathcal{B}}{\textbf{T}(x, y)(1-\mathcal{H}(\phi(x,y)))}}{\sum_{x,y\in\mathcal{B}}(1-\mathcal{H}(\phi(x,y)))}
\end{split}
\label{eq:b}
\end{equation}
where $\mathcal{H}$ is a Heaviside function and computed as
\begin{equation}
H_{\epsilon} ({x}) = \dfrac{1}{2} \left( 1+\frac{2}{\pi}\arctan \left( \frac{{x}}{\epsilon}\right)\right)
\end{equation}

The signed distance function (SDF) \cite{NB_1995} is applied on $\textbf{P}$ to obtain $\phi$.
From Sec.\ref{subsec:offsetcurves}, theory of parallel curve also proved that both $\Gamma^{-\mathcal{B}}$ and $\Gamma^{+\mathcal{B}}$ curves are continuously differential and do not exhibit singularities even if the case band width is larger than the curve’s radius

\textbf{Lower feature loss}: The lower feature loss $\mathcal{L}_3$ is based on minimizing the length of the contour and defined by applying gradient operator $\bigtriangledown$ on $\phi$ as follows:

\begin{equation}
\small
\begin{split}
\centering
    \mathcal{L}_{3} = \sum_{x,y\in \omega}{|\bigtriangledown \phi(x,y)|}
\end{split}    
\end{equation}

Our proposed OsC loss is defined as $\mathcal{L} = \alpha \mathcal{L}_1 + \beta \mathcal{L}_2 + \eta \mathcal{L}_3$. Following similar suggested choice in \cite{chan2001active}, we selected values of {$\alpha, \beta,\eta$} as {0.5, 0.3, 0.2}. Fig.\ref{fig:flowchart} shows an illustration of our proposed OsC loss with Unet architecture as an instance.

\subsection{CNN Architecture}
In this work, we use Unet \cite{Unet} and FCN \cite{long2015fully} architectures as our base segmentation frameworks to evaluate our proposed OsC loss function performance. In Unet, each layer in the down-sampling path consists of two $3 \times 3$ convolution layers, one batch normalization (BN), one rectified linear unit (ReLU) and one max pooling layer. In up-sampling path, bilinear interpolation is used to up-sample the feature maps. In FCN framework, we choose FCN-32 which produces the segmentation map from conv1, conv3, conv7 by using a bilinear interpolation. At down-sampling path, each layer in FCN is designed as same as layer in Unet. To train our model, we employed Adam optimizer, with a learning rate of 1e-2 and a batch size equal to 8.

\begin{table*}[!htb]
  \begin{floatrow}[2]
    \ttabbox%
    {\begin{tabular}{lllll} \hline
    		Losses & DSC & Jac & Pre & Rec \\ \hline\hline
    		\Xhline{2\arrayrulewidth}
    		CE             & 74.47   & 76.01      & 72.00          & 75.00     \\ \cline{1-5}    
        	Dice            & 77.5      & 79.79   & 77.00          & 78.00          \\ \cline{1-5}
		    Focal              & 74.2 & 76.93 & 67.00 &\textbf{83.00   }    \\ \cline{1-5}
		   \textbf{Ours} & \textbf{78.5} & \textbf{80.1}& \textbf{79.00}& 78.00\\
    		\Xhline{2\arrayrulewidth}
    	\end{tabular}}
        {\caption{Comparison on \textbf{DRIVE} dataset and \textbf{FCN}}
          \label{tab:val1}}
        \hfill%
        \ttabbox%
        {\begin{tabular}{lllll}
        \hline
         Losses & DSC & Jac & Pre & Rec \\ \hline \hline
        \Xhline{2\arrayrulewidth} 
        CE      & 77.3 & 73.1 &  76.66   &  79.00 \\  \cline{1-5} 
        Dice    &  76.31 & 71.84 & 75.00 & 76.33  \\  \cline{1-5} Focal  & 72.45  & 67.76 & 68.00 & 77.33    \\  \cline{1-5} 
            \textbf{Ours}  & \textbf{78.46 } &   \textbf{75.24} &  \textbf{78.67} &  \textbf{79.00} \\  \cline{1-5} 
        \Xhline{2\arrayrulewidth} 
        \end{tabular}}
        {\caption{Comparison on \textbf{BRATS 2018} dataset and \textbf{FCN}}
          \label{tab:val2}}
    \end{floatrow}
\end{table*}

\begin{table*}[!htb]
 \begin{floatrow}[2]
    \ttabbox%
        {\begin{tabular}{lllll} \hline
		Losses & DSC & Jac & Pre & Rec \\ \hline\hline
		\Xhline{2\arrayrulewidth}
		CE             & 77.1    & 78.80      & 75.00          & 80.00     \\ \cline{1-5}    
		Dice            & 78.6      & 79.6    & 77.00          & 80.00          \\ \cline{1-5}
		Focal              & 77.7      &  78.9   & 76.00          & 79.00       \\ \cline{1-5}
		\textbf{Ours} & \textbf{79.3} & \textbf{81.2} & \textbf{78.00} & \textbf{82.00} \\  
		\Xhline{2\arrayrulewidth}
	\end{tabular}}
    {\caption{Comparison on \textbf{DRIVE} dataset and \textbf{Unet}}
      \label{tab:val3}}
    \hfill%
    \ttabbox%
    {\begin{tabular}{lllll}
    \hline
     Losses & DSC & Jac & Pre & Rec \\ \hline \hline
    \Xhline{2\arrayrulewidth} 
    CE               &  78.34 & 73.45& 77.33&  80.00  \\ \cline{1-5} 
    Dice             &  77.51 & 72.24& 76.67&   77.67\\ \cline{1-5} 
    Focal             & 75.78      & 77.00    & 67.33          &\textbf{86.00}   \\ \cline{1-5} 
    \textbf{Ours}   &  \textbf{79.78} & \textbf{80.33} & \textbf{78.33}& 81.00\\ 
    \Xhline{2\arrayrulewidth} 
    \end{tabular}}
    {\caption{Comparison on \textbf{BRATS 2018} dataset and \textbf{Unet}}
     \label{tab:val4}}
  \end{floatrow}
\end{table*}

\begin{figure}[!t]
	\centering
	\includegraphics[width=8cm]{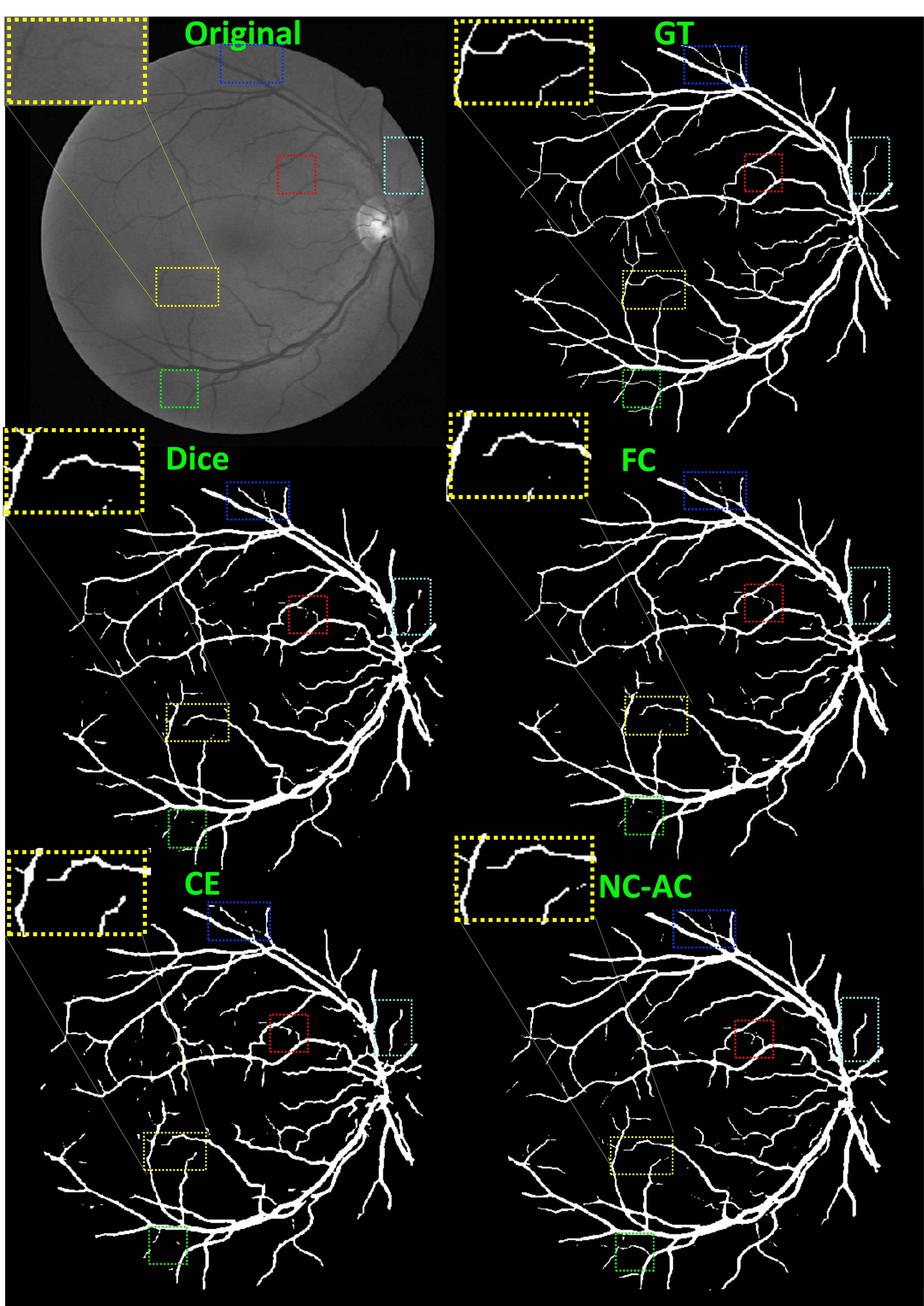}
	\caption{Comparison between our results against other loss functions on 2D Unet framework where the image is from DRIVE dataset. The sub-images in the yellow box is zoomed in}
	\label{fig:res_drive}
\end{figure} 

\begin{figure*}[!t]
	\centering
	\includegraphics[width=18cm]{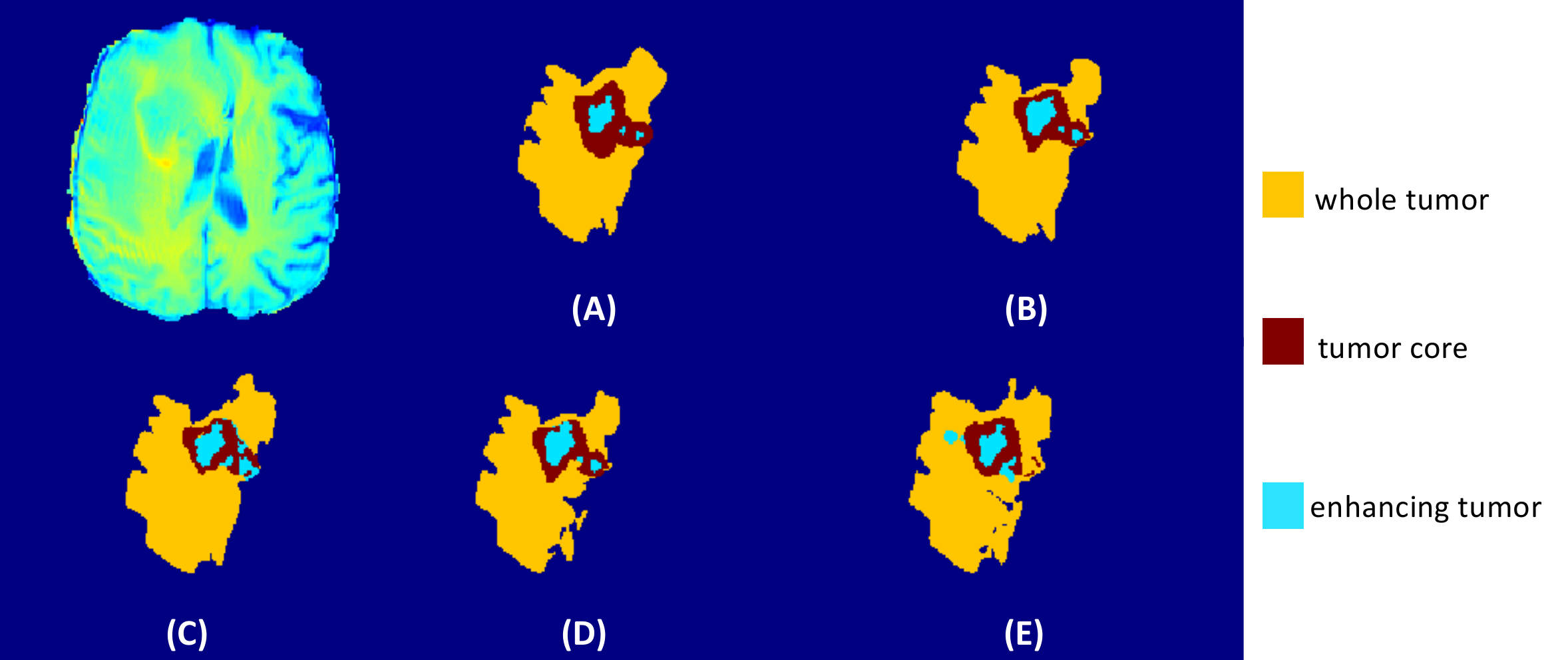}
	\caption{Comparison between our results against other loss functions on 3D Unet framework where the image is from BRATS2018 dataset (A): Groundtruth, (B): OsC loss, (C): CE loss, (D): Dice loss, (E): Focal loss}
	\label{fig:res}
\end{figure*} 

\begin{table*}[]
\begin{tabular}{|l|l|l|l|}
\hline
  Method                    & Patch size & DSC & Hau95 \\ \toprule
NoNew-Net \cite{nonewnet_compare} &   128 $\times$ 128 $\times$ 128   & 84.90     &  5.36       \\ \midrule
Partially reversible \cite{ReversibleNet_compare} & 160 $\times$ 192 $\times$ 160            &  84.39   &   6.33    \\ \midrule
Dual Force \cite{dual_compare}  &     --       &   74.97  &  --     \\ \midrule
Densely-connected CNN \cite{dense_compare}      &    5$\times$ 192 $\times$ 192      &  84.87     &    --   \\ \midrule
OsC + 3D Unet   &  96$\times$ 96$\times$  96      &    84.21      &   6.76       \\ \bottomrule
\end{tabular}
\caption{Comparision between our proposed 3DUnet + OsC loss against the state-of-the-art on validation set of Brats 2018  }
\label{tab:3Dcompare}
\end{table*}

\section{Data \& Experiments}
\subsection{Datasets:}
In the experimental result, we choose both 2D and 3D datasets to evaludate the proposed OsC loss. 

\textbf{2D dataset} The DRIVE \cite{Staal_2014} contains 40 colored fundus photographs, each is sized $565\times 584$. To reduce overfitting problem and to reduce the calculation complexity, our model is trained on small patches sized $48 \times 48 $ which were randomly extracted from the images. The dataset is divided into 20 images for training and validation, 20 images for testing. \\

\textbf{3D dataset} The BRATS 2018 database  \cite{Brats} contains 210 HGG scans and 75 LGG scans while BRATS 2019 contains contains 259 HGG scans and 76 LGG scans. For each scan, there are 4 modalities, i.e., T1, T1C, T2, and Flair which are sampled to an isotropic $1 \times 1 \times 1$ $\text{mm}^3$ resolution and has a dimension of $240 \times 240 \times 155$. In BRATS  databases, there are  three tumor classes: whole tumor (WT), tumor core (TC) and enhanced tumor (ET). For the experiments on validation set, we randomly select 80\% images for training and 20\% for testing on both HGG and LGG scans\\

\textbf{Experimental Results:} For quantitative assessment of the segmentation, the proposed model is evaluated on different metrics, e.g. Dice score (Dice), Jaccard similarity (Jac), Precision(Pre), Recall (or Sensitivity)(Rec), Hausdorff95 (hau95). We have implemented our networks using Torch 1.1.0 and the ADAM optimizer. The experiments are conducted using an Intel CPU, and a NVIDIA TitanX GPU. 

There are 2 main experiments given in this section. In the first experiment, we implement the proposed OsC loss on 2D networks (FCN and Unet) and compare on 2D DRIVE dataset and 2D Brats 2018 when we process 3D volumetric data as a sequence of 2D images. The first experiment demonstrates the effectiveness of proposed OsC loss compared against other common loss functions. Furthermore, we also implement the proposed OsC loss on 3D volumetric data and benchmark against other state-of-the art works on 3D Networks. As for your implementation, we use path size of $64 \times 64 \times 64$. The Adam optimizer with a batch size of two was used to train the network. The initialization learning rate was set as 0.0002 and was decreased ten times every 400 epochs. We trained each model for 1000 epochs. 

The performance of our proposed OsC loss on 2D Unet and 2D FCN architectures on 2D dataset together the comparison against other common loss functions is given in Tables \ref{tab:val1}, \ref{tab:val2}, \ref{tab:val3}, \ref{tab:val4}. Brain tumor segmentation on BRATS 2018 is a 4-class-segment problem and the results are given in Table \ref{tab:val2}, \ref{tab:val4} corresponding to Unet and FCN architecture. Vessel segmentation on DRIVE is a binary segment problem and the results are given in Table \ref{tab:val1}, \ref{tab:val3}.

Table \ref{tab:3Dcompare} shows the performance of our proposed Unet + OsC loss against other state-of-the art network on offline validation set. Compare to NonewNet \cite{nonewnet_compare} and Partically Reversible \cite{ReversibleNet_compare}, our performance is very competitive on Brats 2018 even our input patch size is much smaller. Using smaller patch size help, since our network requires fewer parameters and less memory allocation. 

Fig.\ref{fig:res_drive} visualizes one example of vessel segmentation on DRIVE dataset with 2D Unet architecture on 2D dataset.  Fig.\ref{fig:res} visualizes one example of brain tumor on BRATS 2018 dataset with 3D Unet architecture on 3D dataset. Compared to other loss functions (CE, Dice, Focal), our proposed OsC loss function has achieved the best performance.

\textbf{Discussion:} Our proposed OsC loss function contains three fitting terms which takes into account not only the areas be segmented but also support geometric information around the boundaries and the boundaries length. Instead of using only the areas belonging to the inside/outside of the segmented object which causes the imbalanced-class data problem, the support information around the boundaries is utilized as the second fitting term aiming at addressing this problem. The experiments also demonstrate that our model of OsC loss function performs significantly better than commonly used loss functions e.g. CE, Dice, Focal. Our OsC is not only able to perform on 2D but also easily extended to 3D volumetric data.

\section{Conclusion}
In this paper, we presented a novel loss which utilizes the merits from both active contour and offset curve theories. Our proposed network targets at addressing the problems of imbalanced-class data and weak boundary object segmentation. The proposed network takes into account both higher level features, i.e. region, intermediate feature, i.e. region around the contour and lower level features, i.e. contour. The experimental results have shown that our proposed OsC loss outperforms other common loss functions and is able to be integrated into both 2D and 3D network architectures.

\end{document}